\documentclass[]{iopart}
\pdfoutput=1
\usepackage{iopams}
\usepackage{amssymb}
\usepackage{graphicx}
\usepackage{dcolumn}
\usepackage{bm}
\usepackage{amsfonts}

\begin{document}


\title[Energy Transfer Performance of Electromagnetically-Coupled Nanoresonators]{On the Energy Transfer Performance of Mechanical Nanoresonators Coupled with Electromagnetic Fields}


\author{Hooman Javaheri}
\ead{hooman@ccs.neu.edu}
\address{College of Computer and Information Science, Northeastern University, Boston,  MA 02115, USA}

\author{Bernardo Barbiellini}
\ead{bba@neu.edu}
\address{Physics Department, Northeastern University, Boston, MA 02115, USA}

\author{Guevara Noubir}
\ead{noubir@ccs.neu.edu}
\address{College of Computer and Information Science, Northeastern University, Boston, MA 02115, USA}

\date{\today}


\begin{abstract}
We study the energy transfer performance in electrically and magnetically coupled mechanical nanoresonators. Using the resonant scattering theory, we show that magnetically coupled resonators can achieve the same energy transfer performance as for their electrically coupled counterparts, 
or even outperform them within the scale of interest.  Magnetic and electric coupling are compared in the \emph{Nanotube Radio}, a realistic example of a nano-scale mechanical resonator. The energy transfer performance is also discussed for 
a newly proposed bio-nanoresonator composed of a magnetosomes coated with a net of protein fibers.
\end{abstract}



\section{Introduction}

Mechanical nanoresonators exhibit resonance behaviour involving the mechanical vibrations of the system elements. The natural frequencies of such resonances will, generally, be in the radio frequency range. Nano-scale mechanical resonators coupled with electromagnetic fields have been receiving significant attention recently~\cite{Jensen:2007p4468, Degen:2009p5436, Dykman:2010p4815}. The ability to interact with electromagnetic fields allow such resonators to be essential parts of nano-scale systems. Imaging, sensing, and targeted actuation in nano scale are among several emerging technologies that rely on efficient energy and information transfer. 

In principle, nanoresonators may couple to electromagnetic fields by the charge distributions (Electric coupling) or by the magnetic moment they carry (Magnetic coupling). 
Traditionally, the absorption of waves via the interaction of the electric field with the dipole moment charge distribution of the nanoresonator
has received more attention since many materials are essentially transparent to the magnetic field. 
On the other hand, magnetic coupling of mechanical resonators with electromagnetic 
waves becomes more practical as the size of the system decreases. 
In fact, magnetic coupling holds important advantages over electric coupling. First, magnetically coupled systems can provide more \emph{selective} and \emph{localized} energy transfer. 
That is due to the fact that magnetic fields, unlike electric fields, couple weakly with non-targeted surrounding media, which are often not magnetic. Therefore, magnetic signals penetrate deep in various kinds of media, 
suffer from considerably less attenuations, and interact with targeted resonator inaccessible to electric signals with the same level of energy. In addition, magnetic dipoles are normally more stable than electric dipoles, 
and do not require significant energy from outside to maintain their state. 

This work revisits the interactions of radio-frequency electromagnetic fields with mechanical nanoresonators. In particular, we are interested in quantitative assessment of the energy transfer in such nanoresonators. We use the same methodology presented by Hamam et al.~\cite{Hamam:2007p2142} and focus on low-dissipation conditions that permit resonance. The feasibility of achieving such conditions has been demonstrated in the literature~\cite{Jensen:2007p4468, Stipe:2001p4953}. The outline of this paper is as follows. We first present a general model for mechanical nanoresonators including electric and magnetic coupling mechanisms, and describe the dynamics of the model. Then, we compare the resonant energy transfer performance of the resonator for electric and magnetic coupling using resonant scattering theory. Finally, we sketch a roadmap for a new nanoresonator composed of a magnetite nanoparticle embedded in a net of protein fibers.

\section{Theoretical Model}

\begin{figure}
 \centering
 \includegraphics[width=3.7in]{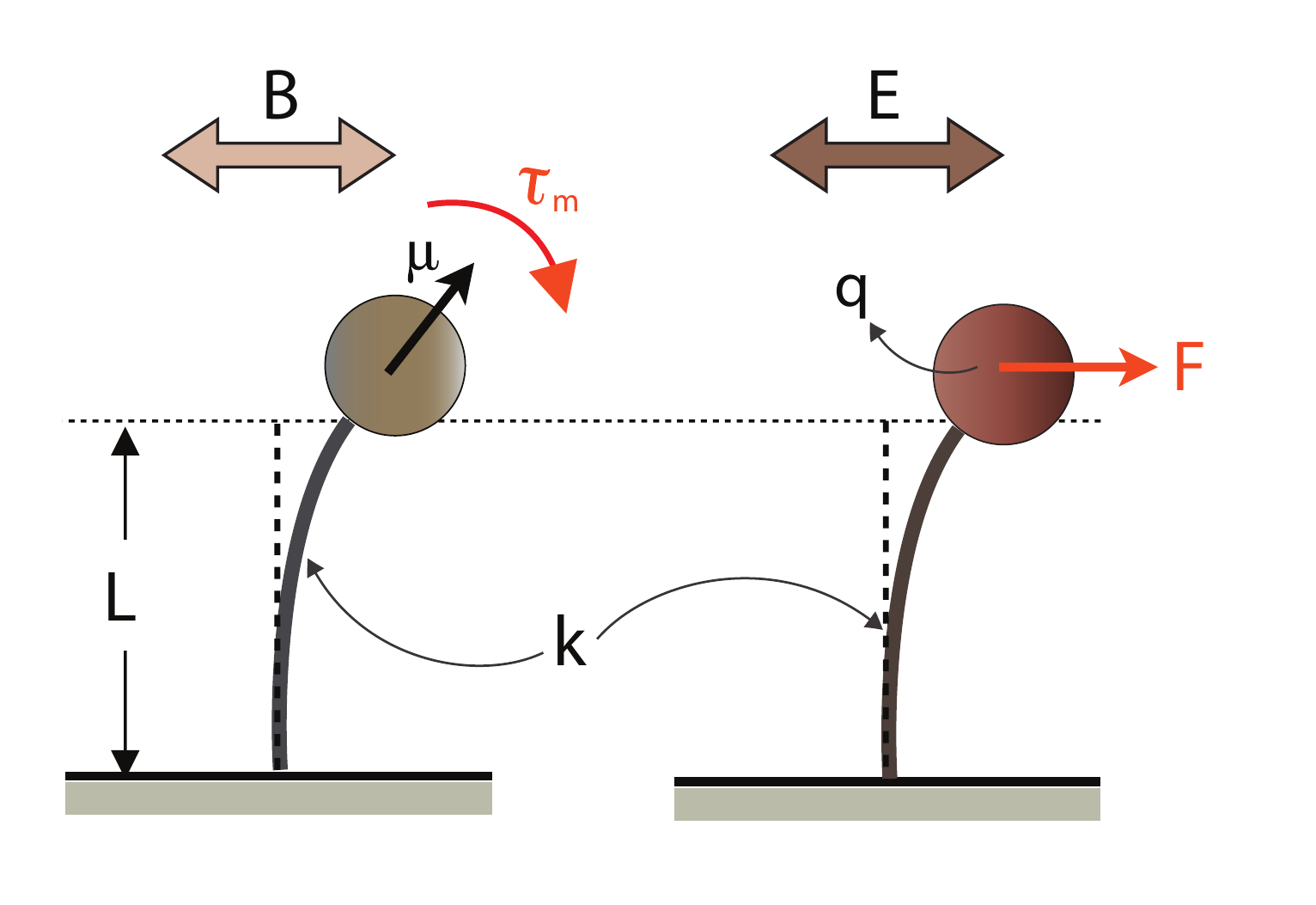}%
 \caption{\label{fig:model} An overview of nanoresonators with electric (right) and magnetic (left) coupling. The viscoelastic properties of the resonators are identical.}
 \end{figure}

In general, the mechanical structure of a nanoresonator consists of an elastic cantilever beam equipped with a specialized tip, which is responsible for electromagnetic interaction, vibrating in a low-viscosity fluid such as low-pressure air. The viscoelastic model of the nanoresonator includes the coefficient of mechanical elasticity, $k$, and the dissipation coefficient, $D$. For a cylindrical beam with a spherical tip, $k \sim EI_{\rm c}/L^3$, where $E$, $I_{\rm c}$ and $L$ are Young's modulus, second moment of cross-section and the length of the beam, respectively. Moreover, as shown in Ref.~\cite{Sazonova:2006p4813}, the combination of intrinsic (e.g., plastic deformation, surface effects) and extrinsic (e.g., viscous forces of the surrounding fluid) dissipation mechanisms determines the value of $D$. Because the size of the nanoresonator is much smaller than the wavelength of the external field, the energy transfer is in the form of interactions between the incoming field and dipole moment of the nanoresonator's tip. As shown in \Fref{fig:model}, we consider two nanoresonators that have identical mechanical structures, yet interact with electromagnetic fields via different coupling mechanisms: Electric coupling $\mathfrak{(E)}$ and \emph{Magnetic coupling $\mathfrak{(M)}$}. In the case of electric coupling, an AC electric field, $\bm E = E\cos(\omega t)$, produces a force, $\bm F$, on an electric charge distribution, $q$, placed at the tip of the nanoresonator and causes oscillatory deflections in the cantilever. A very similar model has been discussed in the \emph{Nanotube Radio}~\cite{Jensen:2007p4468}. For the magnetic coupling, assume the tip of the resonator is made of a ferromagnetic material such as magnetite (Fe$_3$O$_4$) and has a magnetic moment of $\bm\mu$. An AC magnetic field, $\bm B = B\cos(\omega t)$, generates a magnetic torque, ${\cal{T}}_{\rm m} = \bm\mu \times \bm B$, and rotates the tip leading to oscillatory beam deflections. A similar device has been built and used for \emph{ultra-sensitive magnetic resonance force microscopy (MRFM)}~\cite{Degen:2009p5436,Stipe:2001p4953}. 

The dynamics of the system can be expressed by a Langevin equation for the resonator tip. After linearization for small deflections, we have
\begin{equation}
m {\ddot x} + D{\dot x} + kx = F\cos(\omega t) + N(\omega t),
\label{eq:lang}
\end{equation}
where $x$ is the displacement at the tip of the beam, $m$ is the effective mass of the system, and $D$ is the dissipation coefficient. $F = q E$ for electric coupling, while $F = {\cal{T}} / L = \mu B / L$ for magnetic coupling. The term $N(\omega t)$ is an stochastic force with the correlation of $\langle N(t)N(t+\Delta t) \rangle = 2 D k_{\rm B} T \delta(\Delta t)$, where $k_{\rm B}$ and $T$ are the Boltzmann constant and temperature in Kelvin, respectively. In the systems we will consider, the amount of energy stored in the resonator is well above $k_{\rm B} T$. Therefore, we can omit the stochastic term from \eref{eq:lang}. 
The system's natural frequency and the quality factor are given by
\begin{eqnarray}
\omega_0 = \sqrt{\frac{k}{m}} \\
Q  = \frac{\sqrt{k m}}{D}, \label{eq:q}
\end{eqnarray}
while the steady state solution of the system is
\begin{equation}
x(\omega t) = x_{\rm m} \cos(\omega t + \varphi), \label{first}
\end{equation}
where $x_{\rm m}$ and $\varphi$ are the maximum deflection of the tip and the phase shift given by
\begin{eqnarray}
x_{\rm m} (\omega ) = \frac{F / m }{\sqrt{(\omega^2 - \omega_0^2)^2 + (\omega\omega_0 / Q)^2}} \\
\varphi(\omega)  =  \arctan \left(\frac{\omega\omega_0 / Q}{\omega^2-\omega_0^2} \right).
\end{eqnarray}
A resonance can be achieved if $D<\sqrt{2 k m}$. 

The dynamics of the system can also be expressed by the following Langevin equation for rotational oscillation~\cite{Javaheri:2011p6476}.
\begin{equation}
{I \ddot \theta} + C{\dot \theta}  + \kappa\theta = {\cal{T}} \cos(\omega t) + \psi(\omega t)
\end{equation} 
Here $\theta = x / L $ is the angular displacement, $I \sim m L^2$ is the system's second moment of inertia, $\kappa \sim k L^2$ is the rotational spring constant of the cantilever, and $\psi$ is the stochastic torque caused by the thermal noise. For the magnetic coupling, ${\cal{T}} = \mu B$, while ${\cal{T}} = q E L $ in the case of the electric coupling.

\subsection{Energy Analysis} 
We now consider the total energy of the oscillator
\begin{equation}
U = \frac{1}{2}mv^2 + \frac{1}{2}kx^2.
\end{equation}
When $\omega = \omega_0$, this quantity is time-independent and is given by
\begin{equation}
U = \frac{F^2 Q^2}{2 k}.
\end{equation}
One can think of $U$ as the energy capacity of the resonator. An important observation is that $U$ scales with $Q^2$. 

Next, we consider the energy absorbed by the nanoresonator during the relaxation time $\tau=Q/\omega_{\rm r}$. This quantity can be 
calculated by averaging the instantaneous power absorbed by the nanoresonator, $P$, over $\tau$. For our system, $P$  can be written as inner product of incident force and velocity of the resonator
\begin{equation}
P = \bm{F} \cdot \bm{v} = F \cos{(\omega t)} \frac{\rmd x(\omega t)}{\rmd t}.
\end{equation}
Plugging the solution from \eref{first} results in
\begin{eqnarray}
P  &= - F\omega x_{\rm m} \sin(\omega t + \varphi) \cos(\omega t) \nonumber \\
 &=  \frac{-F \omega x_{\rm m}}{2} \times 
 \left( \sin(2\omega t) \cos(\varphi) + (1+ \cos(2 \omega t)) \sin(\varphi) \right) \label{eq:rhs}
\end{eqnarray}
The average of sinusoidal terms in the right hand side of the \eref{eq:rhs} over an integral number of cycles equals zero. Therefore, the average absorbed power, $\bar{P}$, is given by
\begin{equation}
\bar{P} = \frac{F \omega x_{\rm m} \sin(\varphi)}{2}.  
\end{equation} 
At the resonance frequency, $x_{\rm m}=F Q / m \omega_{\rm r}^2$ and $\varphi~=~\pi / 2$, which gives the average absorbed power of $\bar{P} = F^2 Q / 2 m \omega_{\rm r} $.  Thus, the energy deposited in the nanoresonator during the relaxation time $\tau$ is  
\begin{equation}
\Delta U_{\rm r} = \bar{P}\tau = \frac{F^2 Q^2}{2 m \omega_{\rm r}^2} = \frac{F^2 Q^2}{2k}. \label{eq:ding}
\end{equation}
Note that the energy absorbed by the resonator over the relaxation time matches the resonator energy capacity. 
In general, the calculation of the force (or torque) exerted on the nanoresonator through electromagnetic coupling is not straight-forward. 
As an alternative approach, one can use scattering theory~\cite{Hamam:2007p2142, MBlatt:1954p5063}, which allows to work with fluxes instead of forces, to estimate 
the energy deposited on the resonant system. In the next section, we will use this more convenient method to study the resonant energy transfer.

\subsection{Resonant Scattering Analysis}
The coupling between external fields and the nanoresonator consists of an absorption and a scattering process. According to the scattering theory, the power absorbed by the resonant system equals to $P_{\rm a} = \Phi \sigma_{\rm a}$, where $\Phi$ is the incident electromagnetic power flux, and $\sigma_{\rm a}$ is the absorption cross-section given by ~\cite{Hamam:2007p2142, MBlatt:1954p5063}
\begin{equation}
\sigma_{\rm a} (\omega) = 12\pi(\frac{ c}{\omega})^2\frac{\Gamma_{\rm a}\Gamma_{\rm s}}{(\omega - \omega_{\rm r})^2 + (\Gamma_{\rm a} + \Gamma_{\rm s})^2/4}.
\end{equation} 
Here $c$ is the speed of light, $\Gamma_{\rm a}$ is absorption width, and $\Gamma_{\rm s}$ is the scattering width. The widths are the ratio of the power loss to the characteristic energy of the corresponding process. For process $i$, $ \Gamma_i = 1/\tau_i  = \omega /Q_i  $, where $\tau_i$ and $Q_i$ represent the relaxation time and the quality factor, respectively. The total energy absorbed by the resonant system during the resonant process is given by
\begin{equation}
\Delta U (\omega)= P_{\rm a}(\omega)\times \tau_{\rm a} = \frac{\Phi \sigma_{\rm a} (\omega)}{\Gamma} , 
\end{equation}
where $\Gamma = \Gamma_{\rm a} + \Gamma_{\rm s} $ is the total width of the system. For nano-scale systems of interest, $\Gamma \approx \Gamma_{\rm a} $ because $\Gamma_{\rm s} \ll \Gamma_{\rm a}$. 
The maximal energy transfer occurs at the resonant frequency and can be written as
\begin{equation}
{\Delta U}_{\rm r} = \frac{48 \Phi c^2 }{\omega_{\rm r}^2 }\left(\frac{\Gamma_{\rm s}}{\Gamma_{\rm a}^2}\right) = \frac{48 \Phi c^2 }{\omega_{\rm r}^4}~\Gamma_{\rm s} Q_{\rm a}^2 . \label{eq:deltaE}
\end{equation}
$Q_{\rm a}$ is obtained from \eref{eq:q}. 

By definition, the width of the scattering process is equal to the inverse of the decay time of radiating dipole given by 
\begin{equation}
\Gamma_{\rm s} = - \frac{\rmd \ln U(t)}{\rmd t} = - \frac{\rmd U/\rmd t}{U}
\end{equation} 
where $U$ is the energy of the resonator, and $P_{\rm r} = \rmd U/\rmd t$ is the radiative power of the resonator's dipole. For the electric dipole model ($\mathfrak{E}$), we have
\begin{eqnarray}
P_{\rm r}^\mathfrak{E} = -\frac{1}{4 \pi \epsilon_0} \frac{p_0^2 \omega^4}{3 c^3} \label{eq:pr} \\
U^\mathfrak{E}   = \frac{1}{2} k x_{\rm m}^2, 
\end{eqnarray}
where $p_0 = q x_{\rm m}$ is the maximal amplitude for the electric moment of the resonator. Thus one obtains the following scattering width for the system
\begin{equation}
\Gamma_{\rm s}^\mathfrak{E} = \frac{q^2}{4 \pi \epsilon_0} \frac{2\omega^4}{3c^3 k}. \label{eq:gamma_S:e}
\end{equation}
Replacing \eref{eq:gamma_S:e} in \eref{eq:deltaE} results in 
\begin{equation}
{\Delta U}^\mathfrak{E}_{\rm r} = \frac{q^2}{4 \pi \epsilon_0} \frac{ 32 \Phi {Q_{\rm a}}^2}{c k}, \label{eq:deltau:e}
\end{equation}
which confirms that the energy deposited scales as $Q^2/k$. 

Using classical electrodynamics~\cite{jackson1967classical} one can show that the the radiative power of an oscillating magnetic dipole of moment $\mu_{\rm eff}$ 
is given by replacing $p_0$ by $\mu_{\rm eff} /c$ in \eref{eq:pr}. 
In the case of the spherical MNP shown in Fig.1, if $\theta_{\rm m}$ is the maximum angular deflection, then the oscillating part of the magnetic 
dipole of moment is $\mu_{\rm eff} = \mu \sin^2(\theta_{\rm m})$. Therefore, for the magnetic coupling model ($\mathfrak{M}$), 
one can follow the same derivation as in the electric coupling scheme ($\mathfrak{E}$) and obtain
\begin{eqnarray}
P_{\rm r}^\mathfrak{M} = -\frac{1}{4 \pi \epsilon_0} \frac{\mu^2 \omega^4 \theta_{\rm m}^2}{3 c^5}, \label{eq:pr:m} \\
U^\mathfrak{M}   = \frac{1}{2} \kappa \theta_{\rm m}^2 = \frac{1}{2} k L^2 \theta_{\rm m}^2. 
\end{eqnarray}
Therefore, the corresponding radiation width and deposited energy are
\begin{eqnarray}
\Gamma_{\rm s}^\mathfrak{M}  = \frac{\mu^2}{4 \pi \epsilon_0} \frac{2\omega^4}{3c^3 k (Lc)^2}  \label{eq:gamma_S:m} \\
{\Delta U}^\mathfrak{M}_{\rm r}  = \frac{\mu^2}{4 \pi \epsilon_0} \left[ \frac{1}{L c} \right]^2 \frac{ 32 \Phi {Q_{\rm a}^2}}{c k}. \label{eq:deltau:m}
\end{eqnarray}
Note that $Q_{\rm a}$ and $k$ in Equations \eref{eq:deltau:e} and \eref{eq:deltau:m} only depend on the viscoelastic structure of the resonator, 
and are independent from the coupling type (magnetic or electric). Given a similar viscoelastic structure, the energy absorption value for electric and magnetic coupling 
will be comparable if $\mu / Lc \approx q$. For nano-scale systems of interest, the condition $Lc<1000$~m$^2$/s normally holds. 
By comparing  \eref{eq:deltau:m} to \eref{eq:ding}, it is possible to derive an expression for the average magnetic force experienced 
by the nanoresonator over the resonance relaxation time, which is given by 
\begin{equation}
\bar{F} = \sqrt{\frac{\mu^2}{4\pi \epsilon_0} \frac{32 \Phi}{c^3}}. \label{eq:force}
\end{equation}

\section{Applications}
Having discussed the mechanical dynamics of the nanoresonator as well as theoretical formulation for energy transfer performance of different coupling mechanism, we apply our analysis to a possible nanoresonator
sketched in Fig. 2. and we also discuss the feasibility of a bio-nanoresonator composed of protein coated Fe$_{3}$O$_4$ nanoparticles. 
 
In our first example, we compare the energy transfer performance of the magnetic and electric coupling in the \emph{Nanotube Radio}, a realistic example of a mechanical nanoresonator~\cite{Jensen:2007p4468}. 
We replace the electric dipole of the nanotube tip with a magnetic dipole in the form of spherical magnetite nanoparticle. According to the original study, a Nanotube Radio built from a cylindrical carbon nanotube of length  $L \approx 1$ $\mu$m holding a net charge of $q = 200$e absorbs an amount of energy enough to detect radio signals from the electromagnetic radiation. 
To achieve the same amount of energy deposit, the magnetic moment of the replacement tip should be in the order of $\mu \approx qLc=9.6~\times 10^{-15} $~Am$^2$, which can be obtained 
by placing a magnetite nanoparticle of radius $R \sim 160$ nm.

\begin{figure}
 \centering
 \includegraphics[width=3.7in]{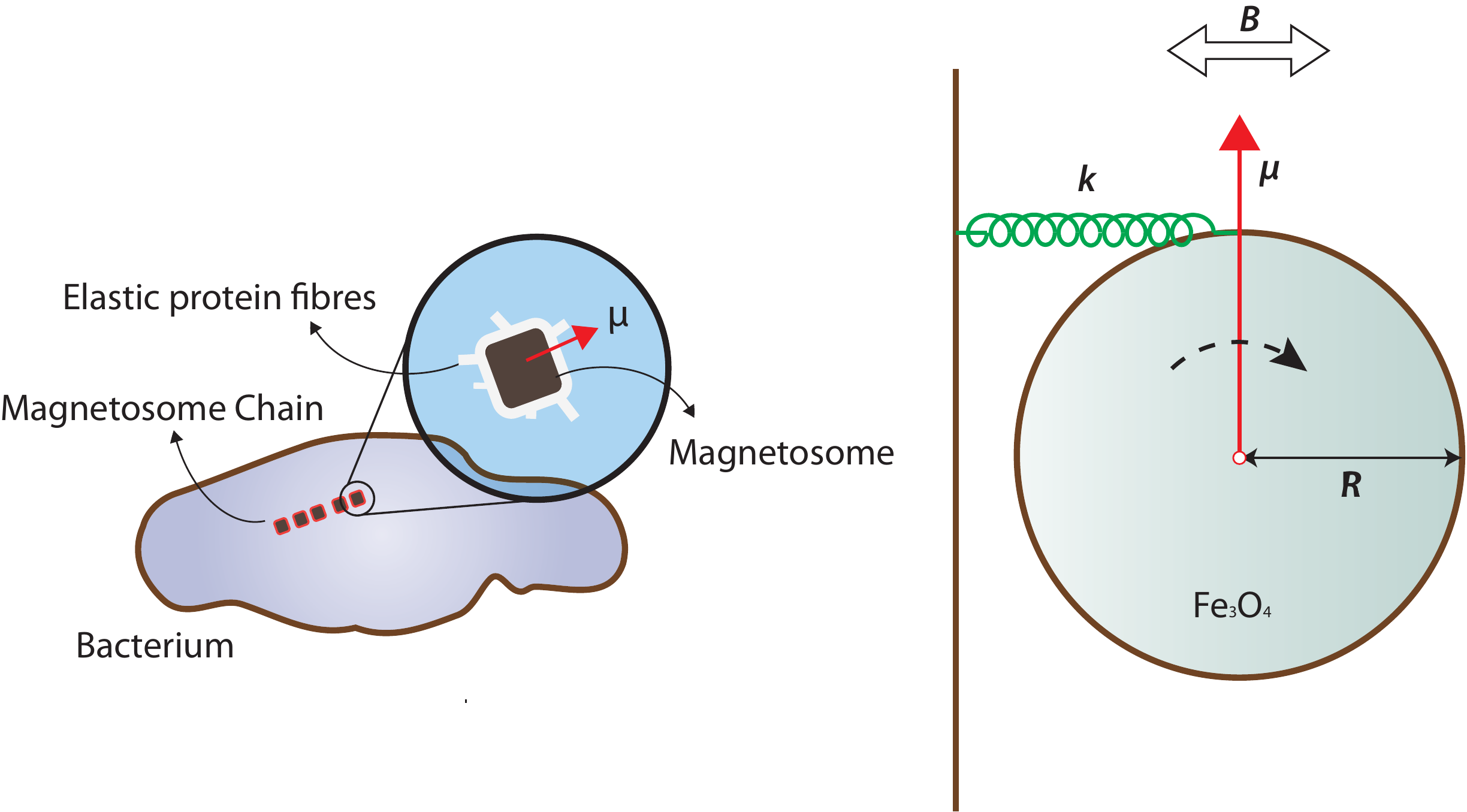}%
 \caption{\label{fig:magnetosome} Magnetosome arrangement in magnetotactic bateria. The magnified part shows how elastic protein fibres embed magnetite (Fe$_3$O$_4$) crystals in the cytoskeleton. Interaction of the magnetic dipole of the crystal with external fields within its viscoelastic environment can be analyzed by our presented theoretical model as a torsional nanoresonator shown on right hand side. Magnetic torque rotates the MNP around its center of mass. The rotational spring constant is given by $\kappa = k R$, where $k$ is the aggregate rigidity of the connecting protein fibers and $R$ is the radius of the MNP. Since the Reynold number of the MNP is very small, the drag forces are given by Stoke's law. Therefore, the rotational damping coefficient is $C=6 \pi \eta R^3$, where $\eta$ is viscosity of the surrounding fluid.}
 \end{figure}

\begin{figure}
      \centering
      \includegraphics[width=5.5in]{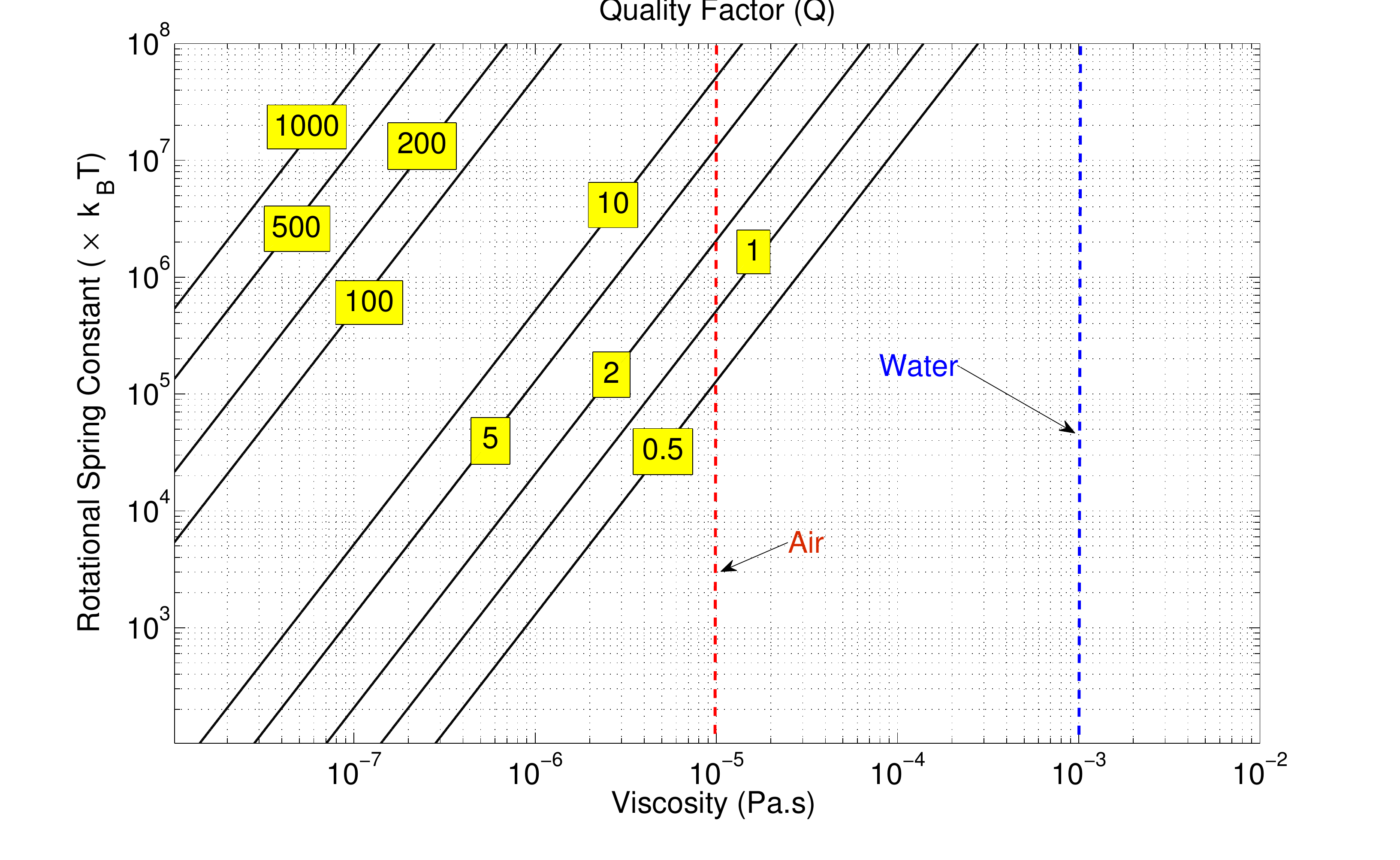}
      \caption{(Color online) Quality factor of the resonance for reasonable range of values for the environment viscosity and the rotational spring constant of elastic environment (in terms of $k_{\rm B}T$). We assume the design includes a magnetite nanoparticle of Radius 100 nm. Note that resonance is possible in the region above $Q=0.5$ line. It is shown how resonance of a given quality can be achieved in lower frequency by reducing the viscosity experienced by the resonator.}
      \label{fig:fre-q}
\end{figure}

Another interesting application is the possibility of transmitting energy to magnetic nanoparticles in biological setting. 
Biogenic magnetite nanoparticles called magnetosomes, first discovered in magnetotactic bacteria~\cite{Blakemore24101975},  are also 
found in the brain of many animals and are believed to participate in determining the orientation in several species 
such as migratory birds~\cite{Kirschvink:2010p124}. 
Interestingly, magnetosomes consist of magnetite particles of radius 50--100 nm, and are
embedded in the cytoskeleton bound to a viscoelastic system
formed by a net of protein fibers. Because magnetic nanoparticles of such size are single-domain with high coercivity, the magnetosome can be represented as a torsional nanoresonator with magnetic coupling (See \Fref{fig:magnetosome}).  
According to Winklhofer et al.~\cite{Winklhofer:2010p128}, the
rigidity of the cytoskeleton can be estimated by  $\kappa = 100 k_{\rm B}T $/Rad per connecting filament. 
Thus, for magnetite particles with a density of $ \rho = 5200 $ Kg/m$^3$, a radius of $R =100$ nm and the number of the connecting filaments ranging from 1 to 1000, the natural frequency of the oscillator will fall between 2 and 66 MHz.  
A resonance is in principle not possible if we adopt the standard viscosity of the cytoplasm~\cite{Adair:2002p4949}. However, in an carefully engineered synthetic system,  
one could lower the drag forces in order to achieve resonance and higher quality factor up to $Q=100$. \Fref{fig:fre-q} shows possible quality factor values for a nanoparticle of radius 100 nm, assuming that the viscosity and elastic constants could be controlled.  

In order to achieve higher quality factor, the nanoresonator should experience smaller viscous resistance. For instance, one can reduce drag forces in the system by coating the magnetite nanoparticle with hydrophobic proteins or lipids. In this case, the hydrophobic coat acts as a lubricant~\cite{Phizicky:1995p6479,Samanta:2008p6481}. In a more sophisticated design, a multi-layer shell of hydrophobic proteins may be used to engulf the nanoresonator and repel water molecules~\cite{Hu:2012p6478}. In order to aggressively reduce the rotational friction, the nanoresonator could be packed in an inorganic shell that completely excludes the system from cytoplasm. The elastic protein fibers may be replaced by synthetic nanowires or nanotubes with carefully designed rigidity. For example, Del Barco et al.~\cite{DelBarco:2001p6480} have demonstrated the possibility to have free rotation of magnetic nanoparticle embedded in a solid matrix.


Assuming that high quality factor, $Q=100$, can be achieved, one finds that an AC magnetic field of intensity $B=3.5$ mT, generating an electromagnetic flux of $10$~W/m$^2$, deposits a significant amount of energy $\Delta U=2500$ $k_{\rm B}T$ into the system over the resonance relaxation time. Since this field intensity is well below the coercivity field of the nanoparticle, we neglect the energy losses via magnetic reversal. If this energy were entirely manifested as heat, the temperature of the magnetosome would be increased by $0.5$ degree Celsius during the relaxation time $\tau=0.1$ $\mu$s. 
As shown in \Fref{fig:fre-q}, $Q=10$ corresponds to $\omega_0= 66$ MHz, in air ($\eta = 10^{-5}$ Pa.s), while the same quality factor can be achieved at frequencies as low as about 1 MHz if the viscosity can be reduced by factor of 100 compared to air.  

%
 
\section{Conclusion}
In conclusion, we have shown that carefully engineered magnetically-coupled nanoresonators can match the energy transfer performance of its electrically-coupled counterpart, while providing a more selective and robust interaction in biological environments. We have used a unifying framework of resonant energy transfer for electrically-coupled and magnetically-coupled mechanical nanoresonators and compared the performance for the two couplings. Our analysis suggests that if the interacting electric dipole of a small electrically-coupled resonator is replaced by a magnetic dipole, a comparable amount of energy can still be deposited on the system. We have considered the example of \emph{Nanotube Radio}~\cite{}, and we have shown that the strength of electromagnetic coupling remains the same using a magnetite nanoparticle of radius 160~nm instead of the charged tip. 
We have proposed a new resonator composed of a magnetosomes embedded in a net of protein fibers, and analyzed its energy transfer performance. We have discussed possible pathways to further improve the quality factor of the resonator. While this article focuses on quantitative aspect of energy transfer, our work also opens up new interesting questions on how to use efficient energy channels to transmit information to a nano-scale device or organism. Characterizing the transmission of information and the channel capacity~\cite{Sidles:2009p5859} will be discussed in future studies.

\ack 
We would like to thank Professor Mark Dykman for his important suggestions and enlightening comments. We also thank the participants of \emph{NSF Workshop of Biologically-enabled Wireless Networks} for stimulating discussions.  This work is supported in part by National Science Foundation under Grant No. NSF CNS-1051240 and also the US Department of Energy, Office of Science, Basic Energy Sciences Contracts No. DE-FG02-07ER46352 and No. DE-FG02-08ER46540 (CMSN), and benefited from allocation of computer time at the NERSC and NU-ASCC computation centers.

\section*{References}
\bibliography{refs}

\begin{thebibliography}{10}

\bibitem{Jensen:2007p4468}
K~Jensen, J~Weldon, H~Garcia, and A~Zettl.
\newblock Nanotube radio.
\newblock {\em Nano letters}, 7(11):3508--3511, Jan 2007.

\bibitem{Degen:2009p5436}
C.~L Degen, M~Poggio, H.~J Mamin, C.~T Rettner, and D~Rugar.
\newblock Nanoscale magnetic resonance imaging.
\newblock {\em P Natl Acad Sci Usa}, 106(5):1313--1317, Jan 2009.

\bibitem{Dykman:2010p4815}
M.~I Dykman, M~Khasin, J~Portman, and S.~W Shaw.
\newblock Spectrum of an oscillator with jumping frequency and the interference
  of partial susceptibilities.
\newblock {\em Phys. Rev. Lett.}, 105(23):230601, Jan 2010.

\bibitem{Hamam:2007p2142}
R.E Hamam, A~Karalis, JD~Joannopoulos, and M~Solja{\v c}i{\'c}.
\newblock Coupled-mode theory for general free-space resonant scattering of
  waves.
\newblock {\em Phys. Rev. A}, 75(5):53801, 2007.

\bibitem{Stipe:2001p4953}
B~Stipe, H~Mamin, T~Stowe, T~Kenny, and D~Rugar.
\newblock Magnetic dissipation and fluctuations in individual nanomagnets
  measured by ultrasensitive cantilever magnetometry.
\newblock {\em Phys. Rev. Lett.}, 86(13):2874--2877, Mar 2001.

\bibitem{Sazonova:2006p4813}
V.A Sazonova.
\newblock A tunable carbon nanotube resonator.
\newblock 2006.

\bibitem{Javaheri:2011p6476}
Hooman Javaheri, Bernardo Barbiellini, and Guevara Noubir.
\newblock Efficient magnetic torque transduction in biological environments
  using tunable nanomechanical resonators.
\newblock {\em Conf Proc IEEE Eng Med Biol Soc}, 2011:1863--6, Jan 2011.

\bibitem{MBlatt:1954p5063}
John~M. Blatt and Victor~F. Weisskopf.
\newblock Theoretical nuclear physics.
\newblock page 864, Jan 1954.

\bibitem{jackson1967classical}
JD~Jackson.
\newblock {\em Classical electrodynamics}.
\newblock 1967.

\bibitem{Blakemore24101975}
R~Blakemore.
\newblock Magnetotactic bacteria.
\newblock {\em Science}, 190(4212):377--379, 1975.

\bibitem{Kirschvink:2010p124}
Joseph~L Kirschvink, Michael Winklhofer, and Michael~M Walker.
\newblock Biophysics of magnetic orientation: strengthening the interface
  between theory and experimental design, Jan 2010.

\bibitem{Winklhofer:2010p128}
Michael Winklhofer and Joseph~L Kirschvink.
\newblock A quantitative assessment of torque-transducer models for
  magnetoreception.
\newblock {\em J R Soc Interface}, 7:S273--S289, Jan 2010.

\bibitem{Adair:2002p4949}
Robert~K Adair.
\newblock Vibrational resonances in biological systems at microwave
  frequencies.
\newblock {\em Biophys J}, 82(3):1147--52, Mar 2002.

\bibitem{Samanta:2008p6481}
Bappaditya Samanta, Haoheng Yan, Nicholas~O Fischer, Jing Shi, D.~Joseph Jerry,
  and Vincent~M Rotello.
\newblock Protein-passivated fe3o4 nanoparticles: low toxicity and rapid
  heating for thermal therapy.
\newblock {\em J. Mater. Chem.}, 18(11):1204, Jan 2008.

\bibitem{Phizicky:1995p6479}
E~M Phizicky and S~Fields.
\newblock Protein-protein interactions: methods for detection and analysis.
\newblock {\em Microbiol Rev}, 59(1):94--123, Mar 1995.

\bibitem{Hu:2012p6478}
Xiao Hu, Peggy Cebe, Anthony~S Weiss, Fiorenzo Omenetto, and David~L Kaplan.
\newblock Protein-based composite materials.
\newblock {\em Materials Today}, 15(5):208--215, May 2012.

\bibitem{DelBarco:2001p6480}
E~Del Barco, J~Asenjo, XX~Zhang, R~Pieczynski, A~Julia, J~Tejada, RF~Ziolo,
  D~Fiorani, and AM~Testa.
\newblock Free rotation of magnetic nanoparticles in a solid matrix.
\newblock {\em Chemistry of materials}, 13(5):1487--1490, 2001.

\bibitem{Sidles:2009p5859}
John~A Sidles.
\newblock Spin microscopy's heritage, achievements, and prospects.
\newblock {\em P Natl Acad Sci Usa}, 106(8):2477--8, Feb 2009.

\end{thebibliography}

\end{document}